\documentclass{jpp}
\usepackage{graphicx}
\usepackage{epstopdf, epsfig}
\usepackage{hyperref}

\def\Xint#1{\mathchoice
   {\XXint\displaystyle\textstyle{#1}}%
   {\XXint\textstyle\scriptstyle{#1}}%
   {\XXint\scriptstyle\scriptscriptstyle{#1}}%
   {\XXint\scriptscriptstyle\scriptscriptstyle{#1}}%
   \!\int}
\def\XXint#1#2#3{{\setbox0=\hbox{$#1{#2#3}{\int}$}
     \vcenter{\hbox{$#2#3$}}\kern-.5\wd0}}

\def\dashint{\Xint-}

\shorttitle{On the radio waves propagating in the pulsar magnetosphere}
\shortauthor{A. G. Mikhaylenko, V. S. Beskin and Ya. N. Istomin}

\title{On the thermal effects on radio waves propagating in the pulsar magnetosphere}

\author{A. G. Mikhaylenko\aff{1}, 
V. S. Beskin\aff{1,2}
  \corresp{\email{beskin@lpi.ru}}
  \and Ya. N. Istomin\aff{1,2}}

\affiliation{\aff{1}Moscow Institute of Physics and Technology, Institutsky per. 9, Dolgoprudny, 147000, Russia
\aff{2}P.N.Lebedev Physical Institute,  Leninsky prosp. 53, Moscow, 119991, Russia}

\begin{document}

\maketitle

\begin{abstract}
Thermal effects on the properties of four electromagnetic waves propagating in the pulsar magnetosphere are analyzed. It is shown that thermal effects change only quantitatively the dispersion properties of superluminal ordinary O-mode freely escaping the pulsar magnetosphere; properties of the extraordinary X-mode remain unchanged. As for two subluminal waves propagating along magnetic field lines, for them thermal effects result in essential absorption. However, this attenuation occurs at considerable distances from the neutron star, so there is no doubt in their existence.
\end{abstract}

\section{Introduction}

Unfortunately, it should be recognized that now the theory of the pulsar radio emission is far from the mainstream of the modern astrophysics. This problem was not solved in 70-80-ties, and we have no consistent theory up to now~\citep{M&T, Michel, Mestel, L&K, L&GS}. Moreover, the discussion for some key points has been actually frozen. For this reason, the new generation is not familiar with the very high level of research that has been carried out in this field~\citep{GK71, GZh, Blandford75, SCh75, Blandford76, BB77, HR78, LM79, LMS79, APS83, LMU83, LP87, U87, MG99}. In recent years, the works have been conducted only in the theory of wave propagation~\citep{BArons1986, BCW91, LP98, Han98, LP00, P2001, Dyks2008, AB2010, Wang1, BPh2012, YuMelrose, Wang3, Wang4, HBP2017}, not in the mechanism of radio emission itself.

So, to date, there are several dozen models of the pulsar radio emission of varying degrees of elaboration (see, e.g.,~\citealt{UU88, BGI88, APS90, KMM91, W94, LBM99, GGM2002, PhUSC}), which are little related to each other. For this reason, we are not going here to carry out their detailed analysis. We consider only one controversial issue, which, apparently, finally finds its solution. The point is that in the theory of the pulsar radio emission there was not agreement on such a seemingly obvious question as the number of radio waves propagating outwards from a neutron star. Most authors insisted that there are only three~\citep{BArons1986, L99, U2006, L2008}, i.e., two transverse and one plasma wave, whereas according to~\citet{BGI88, BGI93} (see also~\citealt{L&GS}) there are four. And the works, claiming that the number of modes is only three, appeared until very recently~\citep{Melrose0}.

Actually, this was the result of misunderstanding, since~\citet{BGI88} did not discuss ''the fourth mode'', but another branch of the plasma mode. Indeed, adopting the convention in which the wave modes are defined in the plasma rest frame, we conclude that there are only three wave modes, i.e., two transverse and one plasma mode. But another convention is to interpret as an additional mode any positive-frequency, forward-propagating solution in the rest frame that arises by Lorentz-transforming a negative-frequency or backward-propagating solution in plasma reference frame. Below we use the second convention, which, after the works by~\citet{Melrose1, Melrose2}, should no longer raise objections. 

On the other hand,~\citet{Melrose2} recalled another important problem related to this issue. The point is that in the work by~\citet{BGI88} it was actually assumed that the energy spread of the plasma outflowing along open field lines is small. The same approximation was later used by~\citet{LP98, LP00}. But according to numerous studies on the generation of a secondary electron-positron plasma in the polar regions of a neutron star~\citep{DH82, GI85, AE2002, ML2010, T2010, TH2015}, in the rest system of the plasma, the temperature is to be of the order of the rest mass of the particles. As a result, the dispersion properties of waves propagating in the pulsar magnetosphere can change significantly. The question of how thermal effects affect the dispersion properties of radio waves in the radio pulsar magnetosphere is the main topic of this work.

Thus, in this paper we analyse in detail how the thermal effects can change the dispersion properties of all the four waves propagating outwards in the pulsar magnetosphere. In particular, we show that for superluminal O-mode the hydrodynamical approximation remains good enough even for high temperature $T \sim 1$ MeV. As to two subluminal modes, for them the kinetic effects result in more effective damping. Recall that this not so important because these modes cannot escape the pulsar magnetosphere as at large distance from the neutron star they propagate along magnetic field lines.

\section{Description of a problem}

To begin with, it is necessary to clearly formulate the task that will be discussed, as well as the area of the considered parameters. This, as we shall see, removes a significant number of contentious issues.

First of all, we do not consider the general dispersion properties of electromagnetic waves, but limit ourselves only to frequencies that fall in the observable radio frequency band 100 MHz--10 GHz. This is one of the main differences from the works of~\citet{Melrose1, Melrose2}, who analyzed the dispersion properties of subluminal waves for arbitrary Lorentz factor of the wave phase velocity $\gamma_{\phi}$.  By the way, it was shown that there is a sufficiently large range of values of $\gamma_{\phi}$, in which there is a good agreement with the results of~\citet{BGI88}. As will be shown below, it is this region that corresponds to the observed radio frequencies. 

Second, in what follows we consider the standard parameters of the plasma in the vicinity of the neutron star, where generation of radio emission is supposed to occur. Assuming that magnetic field on the surface of a neutron star is $B_{0} = 10^{12}$ G, we obtain that $B = 10^{9}$--$10^{12}$ G for distances $r$ up to 10 neutron star radii $R$. For such a large magnetic field the gyrofrequency $\omega_{B} = eB/m_{\rm e}c$ is several order of magnitude larger than radio frequency $\omega$. For this reason in what follows we neglect all the small terms $\sim \omega/\omega_{B}$. In other words, we consider the case of infinite external magnetic field $B \rightarrow \infty$. This implies that only the parallel component of the wave electric field can interact with particles. Another consequence of this approximation is that we can restrict ourselves to one-dimensional distribution function of particles $F(u)$ ($\int F(u){\rm d}u = 1$), where $u = \beta\gamma$ is the particle four-velocity, $\beta = v_{\parallel}/c$, and $\gamma = (1-\beta^2)^{-1/2}$. 

Further, the plasma number density can be determined as
\begin{equation}
n_{\rm e} = \lambda n_{\rm GJ} \approx 0.7 \times 10^{12} \, 
\left(\frac{P}{1\,{\rm s}}\right)^{-1}
\left(\frac{B_{0}}{10^{12}\, {\rm G}}\right)
\left(\frac{r}{10 \, R}\right)^{-3}
\left(\frac{\lambda}{10^4}\right) {\rm cm}^{-3}.
\label{ne}  
\end{equation}
Here 
\begin{equation}
n_{\rm GJ} = \frac{\Omega B}{2 \upi c e}
\label{nGJ}  
\end{equation}
is the Goldreich-Julian particle number density (the minimum value which is necessary to screen the longitudinal electric field), 
\begin{equation}
\lambda = \frac{n_{\rm e}}{n_{\rm GJ}} \sim 10^{4} 
\label{lambda}  
\end{equation}
is the pair production multiplicity~\citep{DH82, GI85, AE2002, ML2010, T2010, TH2015}, and $\Omega = 2 \pi/P$ is the neutron star angular velocity. 

As was shown by~\citet{BGI88}, dispersion properties of normal modes in the pulsar magnetosphere significantly depend on the parameter
\begin{equation}
A_{\rm p} = \frac{\omega_{\rm pe}^2 \gamma_{\rm s}}{\omega^2}
\approx 6 \times 10^3 \,\left(\frac{B_{0}}{10^{12}\, {\rm G}}\right)
\left(\frac{r}{10 \, R}\right)^{-3}
\left(\frac{\lambda}{10^4}\right) 
\left(\frac{\gamma_{\rm s}}{100}\right)
\left(\frac{\nu}{1 \,{\rm GHz}}\right)^{-2},
\label{Ap}  
\end{equation}
where $\omega_{\rm pe} = (4 \upi e^2 n_{\rm e}/m_{\rm e})^{1/2}$ is the plasma frequency and $\gamma_{\rm s} \sim 100$ is the bulk Lorentz-factor of the outflowing plasma. Unfortunately,~\citet{Melrose1, Melrose2} did not discuss this subject in detail. In particular, they provide the value for $n_{2}$
\begin{equation}
n_{2} \approx 1 - \frac{1}{2} \, \frac{\omega_{\rm pe}^2}{\gamma_{\rm s} \omega^2} \, \theta_{\rm b}^2, 
\label{n2}  
\end{equation}
which is valid for $A_{\rm p} <  1$ only. Indeed, for low number density $n_{\rm e}$ (when $A_{\rm p} < 1$), i.e., at large enough distances from the neutron star, plasma has little effect on the properties of the transverse waves, so that with good accuracy we can assume that the refractive indices for extraordinary ($j = 1$) and ordinary ($j = 2$) modes $n_{1,2} \approx 1$. This implies that two orthogonal modes propagates rectilinear in the pulsar magnetosphere. 

On the other hand, the most non-trivial properties take place for $A_{\rm p} > 1$, i.e., just in the radio generation domain $r < 30$--$100 \, R$, where the mixing of transverse ordinary mode and longitudinal plasma mode occurs. As a result, there is a significant difference in the refractive index $n_{2}$ from unity. The expressions for the dependence of the refractive indices on the angle $\theta_{\rm b}$ between the wave vector ${\bf k}$ and magnetic field ${\bf B}$ for all four normal waves obtained by~\citet{BGI88}
\begin{eqnarray}
n_1 & = & 1,
\label{nj1} \\
n_2 & \approx & 1 + \frac{\theta_{\rm b}^2}{4}
- \left(<\frac{\omega_{\rm pe}^2}{\gamma^{3}\omega^2}>
+ \frac{\theta_{\rm b}^4}{16}\right)^{1/2},
\label{nj2}
\\
n_3 & \approx & 1 + \frac{\theta_{\rm b}^2}{4}
+ \left(<\frac{\omega_{\rm pe}^2}{\gamma^{3}\omega^2}>
+ \frac{\theta_{\rm b}^4}{16}\right)^{1/2},
\label{nj3} \\
n_4 & \approx & 1 + \frac{\theta_{\rm b}^2}{2},
\label{nj4}
\end{eqnarray}
matched this last case only. Here the brakets $<...>$ denote the averaging on the distribution function $F(u)$: $<...> = \int ... F(u){\rm d}u$. As we see, another key parameter of our problem is
\begin{equation}
a_{\rm p} = <\frac{\omega_{\rm pe}^2}{\gamma^{3}\omega^2}>.
\label{a}
\end{equation}
In particular, the mode $j = 2$ cannot propagate outward for $a_{\rm p} > 1$, i.e., for small enough frequencies $\omega${\footnote{\citet{BGI88} proposed this property to explain the low-frequency cut-off of the pulsar radio emission.}}.

\begin{figure}
  \centerline{\includegraphics[height=5cm,width=10cm]{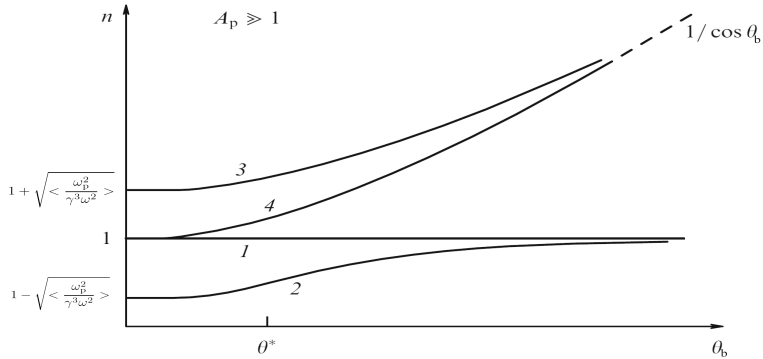}}
  \caption{Normal modes for $A_{\rm p} > 1$ with the refractive indices $n \approx 1$ propagating from the neutron star surface as a function of the angle $\theta_{\rm b}$ between the wave vector ${\bf k}$ and the external magnetic field ${\bf B}$~\citep{BGI88}. Waves $j = 1$ (extraordinary X-mode) and $j = 2$ (ordinary O-mode) can escape the pulsar magnetosphere. Subluminal modes $j = 3, 4$ decay at large angles $\theta_{\rm b}$ and cannot escape the pulsar magnetosphere.}
\label{fig01}
\end{figure}

As we see in Figure~\ref{fig01}, for $A_{\rm p} >  1$ there is a mixing of longitudinal and transverse waves. If for small angles $\theta_{\rm b} < \theta^{\ast}$, where 
\begin{equation}
\theta^{\ast} = <\frac{\omega_{\rm pe}^2}{\gamma^{3}\omega^2}>^{1/4},
\label{theta*}  
\end{equation}
we are dealing with two transverse modes $n_{1} \approx n_{4} \approx 1$ and with two branches of plasma wave $n_{2,3} \approx 1 \pm a_{\rm p}^{1/2}$, then at large angles $\theta_{\rm b} > \theta^{\ast}$ we have two transverse modes with $n_{1} \approx n_{2} \approx 1$ and two Alfv\'en modes with $n_{3} \approx n_{4} \approx 1/\cos\theta_{\rm b}$ propagating along magnetic field lines. As was already noted, these waves are to decay. It should be noted that
\begin{equation}
A_{\rm p} \approx (\theta^{\ast} \gamma_{\rm s})^{4}.
\label{Appp}
\end{equation}
This relation will be used in what follows. Finally, rewriting condition $\theta^{\ast} \gamma_{\rm s} \gg 1$ in the form $\theta^{\ast} \gg \gamma_{\rm s}^{-1}$, we come to another important conclusion that in the domain where $A_{\rm p} \gg 1$ radiation is generated at the angles
$\theta_{\rm b} \ll \theta^{\ast}$.

Once again, we emphasize that the above expressions (\ref{nj1})--(\ref{nj4}) for refractive indices $n_{1, 2, 3, 4}$ were obtained under definite restrictions. In addition to condition $A_{\rm p} > 1$, it was also assumed that the thermal spread over longitudinal momentum $p = m_{\rm e}cu$  is sufficiently small, so that the averaging over the distribution function $F(u)$ does not cover resonant condition $\omega - k_{\parallel}v_{\parallel} = 0$, i.e.,
\begin{equation}
1 - n\beta\cos\theta_{\rm b} = 0.
\label{cheren}  
\end{equation}

This last circumstance was, in fact, a stumbling block. Indeed, exact dispersion equation (for infinite magnetic field) looks like~\citep{BGI88, Melrose2}
\begin{equation}
(1-n^2) \left(1-n^2 +
(1 - n^2 \, \cos^2\theta_{\rm b})
\frac{\omega_{\rm pe}^2}{\omega^2 \, n \cos\theta_{\rm b}}
\int\frac{1}{(1 - n\beta\cos\theta_{\rm b})}\frac{{\rm d}F}{{\rm d}u}{\rm d}u\right) = 0.
\label{DDD}  
\end{equation}
On the other hand, above expressions (\ref{nj1})--(\ref{nj4}) for refractive indices correspond to the cold plasma approximation, when we can take velocity $v$ out of the integration sign. But in reality, the distribution function of secondary particles $F(u)$ is wide enough~\citep{DH82, GI85, AE2002, ML2010, T2010, TH2015}. Therefore, correction for the superluminal mode $j = 2$ and, all the more, the possibility of the damping for subluminal modes $j = 3, 4$ require, of course, a separate detailed consideration~\citep{Melrose1, Melrose2}. However, we recall that the damping of subluminal modes itself, which at large angles propagate along the magnetic field and, therefore, cannot leave the neutron star magnetosphere, has never been questioned~\citep{BGI88}.

Of course, for a quantitative study of thermal effects we must specify the distribution function $F(u)$. As was shown by~\citet{AE2002}, distribution of secondary particles in the plasma rest frame with good accuracy can be approximated by the J{\"u}ttner distribution 
\begin{equation}
F(u^{\prime}) = \frac{\exp(-\rho \gamma^{\prime})}{2K_{1}(\rho)},
\label{Fpp}  
\end{equation}
with $\rho \approx 1$. Here $K_{1}$ is the Macdonald function of order 1. This implies  relativistic temperature $T \approx m_{\rm e}c^2$. Accordingly, in the laboratory (neutron star) reference frame we have{\footnote{In~\citet{Melrose2} the last factor was omitted.}}
\begin{equation}
F(u) = 
%\frac{\exp[-\rho \gamma_{\rm s}\gamma(1 - \beta\beta_{\rm s})]}{2 \gamma_{\rm s}K_{1}(\rho)}.
\frac{\exp[-\rho \gamma_{\rm s}\gamma(1 - \beta\beta_{\rm s})]}{2K_{1}(\rho)}\gamma_{\rm s}(1 - \beta\beta_{\rm s}).
\label{FpJ}  
\end{equation}
However, below we consider a wider class of distribution functions $F(u)$, since the distribution function (\ref{FpJ}) does not always adequately describe the distribution of secondary particles outflowing from the magnetosphere of a neutron star.

Thus, the goal of our study is in clarifying how the thermal effects affect the propagation properties of radio waves in the pulsar magnetosphere. Therefore, in contrast to the works of~\citet{Melrose1, Melrose2}, our main task is in analyzing the dependence of their refractive indices $n$ on the angle $\theta_{\rm b}$. It is clear that in the infinite magnetic field approximation (\ref{DDD}) considered here, the extraordinary X-mode $j = 1$ remains unchanged. Indeed, as electric vector of this wave is perpendicular to external magnetic field, this mode cannot interact with plasma particles. As a result, we have $n_{1} = 1$, i.e., this mode propagates rectilinear in the pulsar magnetosphere. For this reason, below we discuss three other modes only.  

\section{Superluminal O-mode} 

At first, let us consider O-mode $j = 2$; remind that here we consider the case $A_{\rm p} \gg 1$. As shown in Figure~\ref{fig01}, for this wave  $n_{2} < 1$, i.e., it is a superluminal mode. Using now standard Lorentz transformation to plasma rest frame % (see, e.g., ~\citealt{Melrose2})
\begin{equation}
(n\cos\theta)_{\rm rest} = \frac{n\cos\theta - \beta_{\rm s}}{1 - n\beta_{\rm s}\cos\theta},
\label{boost}  
\end{equation}
we obtain for $\gamma_{\rm s} \gg 1$, $A_{\rm p} \gg 1$, and $\theta_{\rm b} = 0$
\begin{equation}
n_{2}^{\rm rest} \approx -1 + \frac{1}{(1-n_{2})\gamma_{\rm s}^2}
\approx -1 + A_{\rm p}^{-1/2}.
\label{boostn2}  
\end{equation}
As $\omega_{\rm rest} = \gamma_{\rm s}\omega(1 - n_{2}\beta) > 0$, one can conclude that for $\theta_{\rm b} = 0$ in the plasma rest frame the O-mode $j = 2$ corresponds to superluminal branch of plasma wave propagating to the star surface.

On the other hand, as for superluminal O-mode the denominator in (\ref{DDD}) is positive for all particle four-velocities $u$, one can obtain integrating by part
\begin{equation}
1-n_{2}^2 - (1 - n_{2}^2 \, \cos^2\theta_{\rm b})
\frac{\omega_{\rm pe}^2}{\omega^2 }
<\frac{1}{\gamma^3(1 - n_{2}\beta\cos\theta_{\rm b})^2}> \, = \, 0.
\label{DDD2}  
\end{equation}
For cold plasma (and for $|n - 1| \gg |\beta - 1|$, which is just true for $A_{\rm p} \gg 1$) one can put $\beta = 1$, and we return to (\ref{nj2}). On the other hand, for wide enough distribution function, i.e., for parameter $\rho \sim 1$ incoming into J{\"u}ttner distribution function (\ref{Fpp}), the value of $\beta$ under the integral sign cannot be considered constant, and therefore a more detailed study is required.

\begin{figure}
\centerline{\includegraphics[height=5cm,width=10cm]{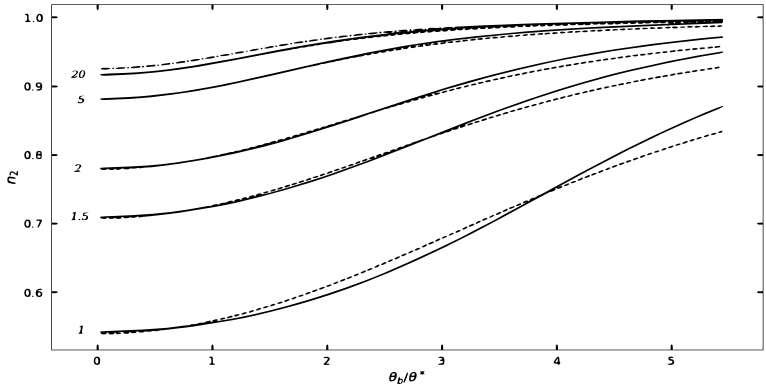}}
  \caption{Dependence of refractive index $n_{2}$ on the angle $\theta_{\rm b}$ for different parameter $\rho$ for $\gamma_{\rm s} = 100$, $\lambda = 10^{4}$, $P = 1$ s, $B = 10^{11}$ G, and $\nu = 1$ GHz ($A_{\rm p} \sim 10^{6}$). Dashed lines correspond to analytical expression (\ref{nj2}) for different parameters $\rho$, and the dash-dotted line corresponds to cold outflow $\rho \rightarrow \infty$.}
\label{fig02}
\end{figure}

In Figure~\ref{fig02} we show how the refractive index $n_{2}$ depends on the angle $\theta_{\rm b}$ for $A_{\rm p} \gg 1$ for different parameters $\rho$ (solid lines). Dashed lines correspond to analytical expression (\ref{nj2}) where the condition $\beta = 1$ was assumed, and the dash-dotted line to cold outflow $\rho \rightarrow \infty$. As we see, even for hot enough plasma (i.e., for $\rho \sim 1$) the analytical evaluation (\ref{nj2}) reproduces good enough the dependence for O-mode refractive index on angle $\theta_{\rm b}$. On the other hand, for $\rho \sim 1$ the difference in the refractive index from the unity is already two times larger than for cold plasma. Note that in this Figure, the angle $\theta_{\rm b}$ is expressed in terms of $\theta^{\ast}$, which is different for different frequencies $\omega$. 
%Nevertheless, qualitatively the form of this dependence remains the same as for cold plasma. 

Remind that it is the wave $j = 2$ that escape from the pulsar magnetosphere as the ordinary O-mode~\citep{BArons1986, BPh2012}. In this case, the difference in the refractive index from unity leads to a deviation of the wave from the magnetic axis. As the result, the mean profile formed by the O-mode turns out to be wider than for the extraordinary X-mode $j = 1$. 
%Since for $\rho = 1$ the difference in the refractive index from unity is twice as large as for the cold plasma, one can conclude that in this case the width of the mean profile will also be twice as large as for a cold plasma. 

To obtain the width of the directivity pattern $W_{\rm r}$, we must now determine the propagation path of the O-mode. As was shown by~\citet{BGI88}, for small angles $\theta_{\perp} = k_{\perp}/k \ll 1$ and $\alpha = (3/2)r_{\perp}/l \ll 1$ where $\alpha$ is the angle between the magnetic field and magnetic axis and  $r_{\perp}$ and $k_{\perp}$ are the components of the vectors perpendicular to magnetic axis, in dipole magnetic field the equations of geometric optics ${\rm d}{\bf r}/{\rm d} t = \partial \omega/\partial {\bf k}$, ${\rm d}{\bf k}/{\rm d} t = -\partial \omega/\partial {\bf r}$ are reduced to the following system
\begin{eqnarray}
\frac{{\rm d}r_{\perp}}{{\rm d} l} & = & \theta_{\perp} - \frac{1}{n} \frac{\partial n}{\partial \theta_{\rm b}}, 
\label{go1} \\
\frac{{\rm d}\theta_{\perp}}{{\rm d} l} & = & \frac{3}{2} \, \frac{1}{nl}
\frac{\partial n}{\partial \theta_{\rm b}}.
\label{go2}
\end{eqnarray}
Here $l$ is the distance from the star center, and we use relations $\theta_{\rm b} = \alpha - \theta_{\perp}$ and ${\rm d}l/{\rm d}t = c/n_{2}$. As a result, we obtain for the derivative 
\begin{eqnarray}
\frac{\partial n_{2}}{\partial \theta_{\rm b}} = \theta_{\rm b}\frac{A}{1 + A},
\end{eqnarray}
where
\begin{eqnarray}
A = 4\frac{\omega_{\rm pe}^2}{\omega^2} \, <\frac{u(2u^2\xi-1)}{(2u^2\xi+1)^3}>,
\end{eqnarray}
and $\xi = 1 - n_{2} - \theta_{\rm b}^2/2$ is to be a root of the dispersion equation
\begin{eqnarray}
\xi - \frac{\theta_{\rm b}^2}{2} = 4\frac{\omega_{\rm pe}^2}{\omega^2} \, <\frac{u\xi}{(2u^2\xi+1)^2}>.
\end{eqnarray}
In the limit $u^2\xi \gg 1$ we reduce to 'hydrodynamical' relations
\begin{eqnarray}
\frac{{\rm d}r_{\perp}}{{\rm d} l} = \theta_{\perp} + \frac{(\alpha - \theta_{\perp})}{2}
\left[1-\frac{(\alpha - \theta_{\perp})^2}{[16 a_{\rm p}(l) +(\alpha - \theta_{\perp})^4]^{1/2}} \right], 
\label{goo1} \\
\frac{{\rm d}\theta_{\perp}}{{\rm d} l} = \frac{3}{4} \, \frac{(\alpha - \theta_{\perp})}{l}
\left[1-\frac{(\alpha - \theta_{\perp})^2}{[16 a_{\rm p}(l) +(\alpha - \theta_{\perp})^4]^{1/2}} \right],
\label{goo2}
\end{eqnarray}
obtained by~\citet{BGI88}.
Here $a_{\rm p}$ is given by relation (\ref{a})
%\begin{equation}
%J_{2} % = \frac{\omega_{\rm pe}^2}{\omega^2} <\frac{(1-n_{2}+\theta_{\rm b}^2/2)^2}{\gamma^3(1-n_{2}+\theta_{\rm b}^2/2+u^{-2}/2)^2}> 
%\,\, \approx \,\, <\frac{\omega_{\rm pe}^2}{\gamma^3\omega^2}>,
%\end{equation}
and $n_{2}$ is to be found as a solution of equation (\ref{DDD2}). Note that during integration it is necessary to take into account the dependence of $a_{\rm p} = <\omega_{\rm pe}^2/(\gamma^3\omega^2)>$ via $\omega_{\rm pe}$ on $l$ (Figure~\ref{fig02} shows the dependence of the refractive index $n_{2}$ on the angle $\theta_{\rm b}$ at a given point, not along the beam propagation).

\begin{figure}
\centerline{\includegraphics[height=5cm,width=10cm]{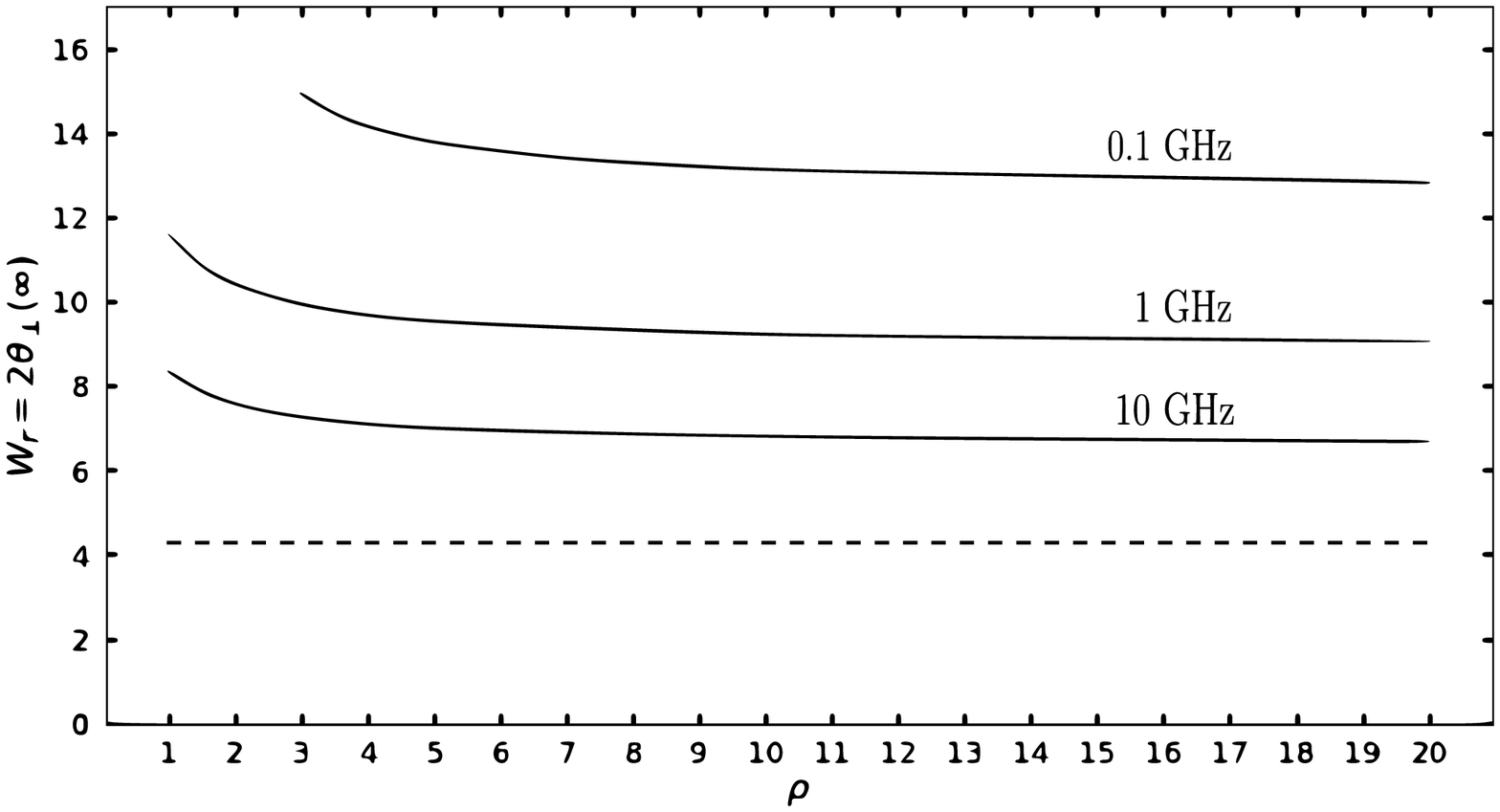}}
  \caption{Dependence of the width of the directivity pattern $W_{\rm r}$ (\ref{Wr}) (in degrees) on parameter $\rho$ for $\gamma_{\rm s} = 100$, $\lambda = 10^{4}$, $P = 1$ s,  $\nu = 1$ GHz, magnetic field on the star surface $B = 10^{12}$ G, and for starting point $l=3\,R$. Dashed line corresponds to rectilinear propagation.}
\label{fig03}
\end{figure}

As a result, integrating the system of equations (\ref{go1})--(\ref{go2}), we obtain for the width of the directivity pattern
\begin{equation}
W_{\rm r} = 2\, \theta_{\perp}(\infty).
\label{Wr}
\end{equation}

Dependence of the width of the directivity pattern $W_{\rm r}$ (\ref{Wr}) on parameter $\rho$ is shown on Figure~\ref{fig03} for the same parameters as in Figure~\ref{fig02}; we start at the point $ l= 3\,R$ where $\theta_{\perp} = \alpha$ (radiation along magnetic field line). Dashed line corresponds to rectilinear propagation when $W_{\rm r} = 2 \, \alpha(3\,R)$. As we see, thermal effects do not significantly affect the width of the directivity pattern. 

Detailed study of the effects of the refraction of an ordinary wave on the formation of mean profiles of radio pulsars will be carried out in a separate work. Nevertheless, we note here that under the condition $\rho \gg 1$, the dependence of the window width $W_{\rm r}$ on the frequency $\nu$ shown in Figure~\ref{fig02} with good accuracy satisfies the dependence 
\begin{equation}
W_{\rm r} \approx 8^{\circ}\nu_{\rm GHz}^{-0.14}
\end{equation}
predicted by~\citet{BGI88} for O-mode for $P = 1$ s, $\lambda = 10^4$, and $r_0 = 3\,R$.

\section{Subluminal  plasma-Alfv{\'e}n mode} 

Now, let us consider mixing plasma-Alfv{\'e}n mode $j = 3$ ($n_{3} > 1$, see Figure~\ref{fig01}); again, we consider the case $A_{\rm p} \gg 1$. For this mode for $\theta_{\rm b} = 0$
\begin{equation}
n_{3}^{\rm rest} \approx -1 + \frac{1}{(1-n_{3})\gamma_{\rm s}^2}
\approx -1 - A_{\rm p}^{-1/2},
\label{boostn2}  
\end{equation}
and $\omega_{\rm rest} = \gamma_{\rm s}\omega(1 - n_{2}\beta) < 0$.
Thus, for $\theta_{\rm b} = 0$ in the plasma rest frame this mode corresponds to subluminal branch of plasma wave also propagating to the star surface. On the other hand, for large enough angles $\theta_{\rm b}$ in the laboratory frame equation (\ref{nj3}) gives \mbox{$n_{3} \approx 1 + \theta_{\rm b}^2/2 \approx 1/\cos\theta_{\rm b}$.} This dispersion equation corresponds to Alfv{\'e}n mode propagating along magnetic field line. 

It is clear that the main difference between subluminal ($j = 3$) and superluminal \mbox{($j = 2$)} wave is that it becomes possible to perform the Cherenkov resonance condition (\ref{cheren}). In this case, since we consider the problem of wave propagation in a curved external magnetic field when the angle $\theta_{\rm b}$ changes along the propagation path, the resonance condition is to be fulfilled even for a cold plasma. That is why in Figure~\ref{fig01} it is the angle $\theta_{\rm b}$ that is chosen as the main parameter.

However, for cold plasma, the resonance condition (\ref{cheren}) occurs for the angles $\theta_{\rm b}$ larger than the angle $\theta^{\ast}$. This result can be obtained immediately if we write down the resonance condition in the form $(n - 1) - 1/2\gamma^2 - \theta_{\rm b}^2/2 = 0$. For $n_{3} > 1 + (\omega_{\rm pe}^2/\gamma^3\omega^2)^{1/2}$ and for $A_{\rm p } \gg 1$ we just obtain  $\theta_{\rm res} > \theta^{\ast}$. Therefore, we can conclude that for cold plasma, the damping of plasma-Alfv{\'e}n mode occurs only for large enough angles $\theta_{\rm b}$ (see Figure~\ref{fig01}).

On the other hand, for a fairly wide distribution function $F(u)$ there are particles that are in resonance with the wave even for $\theta_{\rm b} = 0$. For ultrarelativistic plasma ($\gamma_{\rm s} \gg 1$) and for small angles $\theta_{\rm b} \ll \theta^{\ast}$, the resonance condition can be written in the form $\gamma_{\rm res} \approx 1/\theta^{\ast}$. As a result, using relation (\ref{Appp}), we obtain for the exponent $a = \rho \gamma_{\rm s} \gamma (1 - \beta_{\rm s} \beta)$ in the J{\"u}ttner distribution $F(u) \propto \exp(-a)$
\begin{equation}
a \approx \rho \gamma_{\rm s}\theta^{\ast} \sim \rho A_{\rm p}^{1/4}.
\label{res3}
\end{equation}
Thus, for $A_{\rm p} \gg 1$ the damping of the plasma-Alfv{\'e}n mode for $\theta_{\rm b} = 0$ is to be weak even for $\rho \sim 1$. % Note that for this mode the resonant particles have the energies much smaller than the average energy of outflowing plasma ($\gamma_{\rm res} \ll \gamma_{\rm s}$).

Let us look for the solution of the dispersion equation (\ref{DDD}) in the form
\begin{equation}
n_{3}\cos\theta_{\rm b} = 1 + x.
\label{1j3}
\end{equation}
Then for small angles $\theta_{\rm b} \ll 1$ and relativistic energies ($\beta \approx 1 - 1/2u^2$) we obtain 
\begin{equation}
x\left(x + \frac{\theta_{\rm b}^2}{2}\right) = I_{3}(x),
\label{2j3}
\end{equation}
where
\begin{equation}
I_{3}(x) = \frac{\omega_{\rm pe}^2}{\omega^2(1+x)} \left[ 
\dashint \frac{2u^2 x^2}{(2u^2 x - 1)} \, \frac{{\rm d}F}{{\rm d}u} \, {\rm d}u + i \upi \frac{x^{1/2}}{2\sqrt{2}} \, \frac{{\rm d}F}{{\rm d}u}\bigg|_{u = (2x)^{-1/2}}\right].
\label{3j3}
\end{equation}
Here, as usual, for small imaginary part ${\rm Im}\, n$ in comparison with its real part the real part of the integral $\dashint$ is to be taken in the sense of the principal value. 

As a result, for small angles $\theta_{\rm b} \ll \theta^{\ast}$, when $x \approx I_{3}^{1/2}$,
we have $x u^2 \gg 1$ due to condition $A_{\rm p} \gg 1$. This implies that the resonant particles correspond to the low-energy part of the distribution function ($\gamma_{\rm res} \ll \gamma_{\rm s}$). As a result, for cold outflow we obtain $I_{3} = J_{3}$, where
\begin{equation}
J_{3}(x) = \frac{\omega_{\rm pe}^2}{\omega^2(1+x)} \dashint \frac{2u^2 x^2}{(2u^2 x - 1)} \, \frac{{\rm d}F}{{\rm d}u} \, {\rm d}u  \, \approx \, <\frac{\omega_{\rm pe}^2}{\gamma^3 \omega^2}>,
\label{3j5}
\end{equation}
reproducing asymptotic solution $n_{3}$ (\ref{nj3}).

As we see, the leading (real) term in $I_{3}$  actually does not depend on $x$. Therefore, for weak damping, the imaginary part of $n_{3} \approx 1 + x_{3} + \theta_{\rm b}^2/2$ is completely determined by the second term 
\begin{equation}
{\rm Im} \, n_{3} = \frac{\upi}{2\sqrt{2}} \,
\frac{[(16 \, J_{3} + \theta_{\rm b}^4)^{1/2} - \theta_{\rm b}^2]^{1/2}}{(16 \, J_{3} + \theta_{\rm b}^4)^{1/2}} \, 
\frac{{\rm d}F}{{\rm d}u}\bigg|_{u = (2x)^{-1/2}}.
\label{Imn3}
\end{equation}
Since, as was already emphasized, the resonance condition is achieved in the low-energy region, where ${\rm d}F/{\rm d} u > 0$. Nevertheless, ${\rm Im}\, n_{3} > 0$, which corresponds to attenuation of the wave $j = 3$. This is due to the fact that the wave frequency is negative in the plasma rest reference frame.

\begin{figure}
\centerline{\includegraphics[height=5cm,width=10cm]{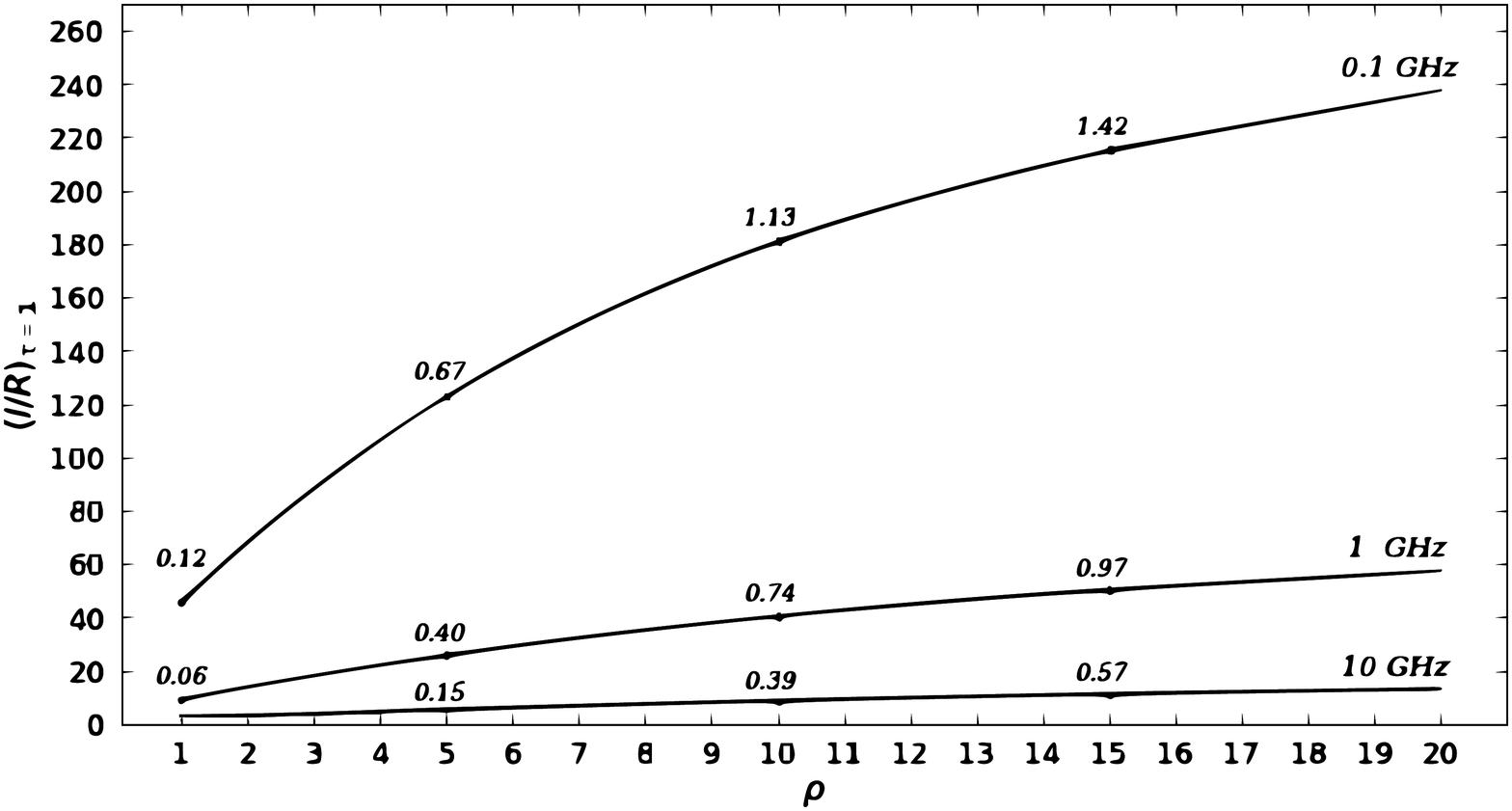}}
  \caption{The levels $l = r_{\rm d}$ at which the optical depth $\tau$ (\ref{tau}) for mode $j = 3$
becomes equal to one for the same parameters as on Figure~\ref{fig03}. The numbers show the ratios $\theta_{\rm b}/\theta^{\ast}$ at the level $\tau = 1$.}
\label{fig04}
\end{figure}

Figure~\ref{fig04} shows the levels $l = r_{\rm d}$ at which the optical depth for the mode $j = 3$
\begin{equation}
\tau = \frac{\omega}{c} \int_{l=r_{0}}^{l=r_{\rm d}} \, {\rm Im} \, n_{3}{\rm d}l
\label{tau}
\end{equation}
becomes equal to unity. Integration was carried out using general relations (\ref{go1})--(\ref{go2}) resulting in~\citep{BGI88} 
\begin{eqnarray}
\frac{{\rm d}r_{\perp}}{{\rm d} l} = \theta_{\perp} + \frac{(\alpha - \theta_{\perp})}{2}
\left[1 + \frac{(\alpha - \theta_{\perp})^2}{[16 a_{\rm p}(l) +(\alpha - \theta_{\perp})^4]^{1/2}} \right], 
\label{gogo1} \\
\frac{{\rm d}\theta_{\perp}}{{\rm d} l} = \frac{3}{4} \, \frac{(\alpha - \theta_{\perp})}{l}
\left[1 + \frac{(\alpha - \theta_{\perp})^2}{[16 a_{\rm p}(l) +(\alpha - \theta_{\perp})^4]^{1/2}} \right].
\label{gogo2}
\end{eqnarray}
As in all previous figures, we put the radiation radius $r_{0} = 3\,R$. The numbers show the current ratios $\theta_{\rm b}/\theta^{\ast}$ at the level $\tau = 1$. 

As we can see, even for $\rho = 1$ the wave damps at the heights significantly higher than the radiation level. On the other hand, for $\rho \sim 1$ the damping occurs at the angles $\theta_{\rm b}$ even smaller than local $\theta^{\ast}(l)$. Wherein, for the parameter area of interest to us ($\rho > 1$, $\nu < 10$ GHz), the second (imaginary) term in (\ref{3j3}) actually turns out to be much less than the firsrt (real) one. 

\section{Subluminal Alfv{\'e}n mode}

Finally, let us consider Alfv{\'e}n mode $j = 4$. For $\theta_{\rm b} = 0$  it corresponds to transverse ordinary O-mode ($n_{4} = 1$), but for large enough $\theta_{\rm b}$ (and again for $A_{\rm p} \gg 1$!) we obtain $n_{4} \approx 1/\cos\theta_{\rm b}$, i.e., this mode also propagates along magnetic field line. Unlike the mode $j = 3$, for mode $j = 4$ the resonance condition cannot be satisfied for $\theta_{\rm b} = 0$. However, it becomes possible for $\theta_{\rm b} > 0$, the resonant particles having the energies much larger than the average energy of outflowing plasma ($\gamma_{\rm res} \gg \gamma_{\rm s}$).

Due to (\ref{nj4}), for this mode it is also convenient to look for the solution of the dispersion equation (\ref{DDD}) in the form
\begin{equation}
n_{4}\cos\theta_{\rm b} = 1 + x.
\label{1j4}
\end{equation}
Then for small angles $\theta_{\rm b} \ll 1$ and relativistic energies ($\beta \approx 1 - 1/2u^2$) we obtain 
\begin{equation}
x = \frac{\theta_{\rm b}^2/2}{I_{4}(x) - 1},
\label{2j4}
\end{equation}
where
\begin{equation}    
I_{4}(x) = - \frac{\omega_{\rm pe}^2}{\omega^2} \left[
\dashint \frac{2u^2}{1 - 2 u^2 x} \, \frac{{\rm d}F}{{\rm d}u} \, {\rm d}u - i \upi \frac{x^{-3/2}}{2\sqrt{2}} \frac{{\rm d}F}{{\rm d}u}\bigg|_{u = (2x)^{-1/2}}\right].
\label{3j4}
\end{equation}
Here, in contrast to (\ref{3j3}), we have $xu^2 \ll 1$ (as was already stressed, this implies that the resonant particles correspond to the high-energy part of the distribution function). As $I_{4}(0) \approx 4 A_{\rm p}$, we can conclude that for $A_{\rm p} \gg 1$ the disturbance $x$ to expression $n_{4}$ (\ref{nj4}) is indeed small, for $\theta_{\rm b} = 0$ resonant particles being absent. Certainly, as before, this expression can only be used if the second term in square brackets is much smaller than the first term.

As a result, for small angles $\theta_{\rm b}$ and for $A_{\rm p} \gg 1$ we have 
\begin{equation}
x_{4} = \frac{\theta_{\rm b}^2/2}{J_{4}},
\label{x4}
\end{equation}
where
\begin{equation}
J_{4} = 4 \, \frac{\omega_{\rm pe}^2<u> }{\omega^2} \, \approx \, 4 \, A_{\rm p}.
\label{J4}
\end{equation}
Thus, here the leading (real) term in $J_{4}$ again does not depend on $x$. Therefore, for
weak damping, the imaginary part of $n_{4} \approx 1 + x_{4} + \theta_{\rm b}^2/2$ is completely determined by the
second term
\begin{equation}
{\rm Im} \, n_{4} = -\frac{\upi}{8} \,
\frac{J_{4}^{1/2}}{<u>\theta_{\rm b}} \, 
\frac{{\rm d}F}{{\rm d}u}\bigg|_{u = (2x)^{-1/2}}.
\label{Imn4}
\end{equation}
Here we neglect the terms of the order of $A_{\rm p}^{-1/2}$. Since now the resonance condition is achieved in the high-energy region, where ${\rm d}F/{\rm d} u < 0$, we again have ${\rm Im}\, n_{4} > 0$ corresponding to attenuation of the wave $j = 4$.

Note that in reality the relation (\ref{2j4}) has no singularity. Indeed, expanding $I_{4}$ as
\begin{equation}
I_{4}(x) = - \frac{\omega_{\rm pe}^2}{\omega^2}
\dashint 2u^2(1 + 2 u^2 x) \, \frac{{\rm d}F}{{\rm d}u} \, {\rm d}u
\approx 4 \frac{\omega_{\rm pe}^2<u>}{\omega^2} + 16 \frac{\omega_{\rm pe}^2<u^3>}{\omega^2}x 
+ ...
\end{equation}
we obtain assuming that $\omega_{\rm pe}^2<u^3>/\omega^2 \approx J_{4}\gamma_{\rm s}^2$
\begin{equation}
x \approx \frac{-(J_{4} - 1) \pm 
[(J_{4} - 1)^2 + 8\,J_{4}\gamma_{\rm s}^2\theta_{\rm b}^2]^{1/2}}{8\,J_{4}\gamma_{\rm s}^2},
\end{equation}
where the sign 'plus' corresponds to $J_{4} > 1$, and the sign 'minus' to $J_{4} < 1$.
As a result, as one can easily check, $n_{4}(\theta_{\rm b})$ remains always larger that unity
if $J_{4} > 1$, and v.v. On the other hand, for $A_{\rm p} \ll 1$ ($I_{4} \ll 1$) expression (\ref{2j4}) results in
\begin{equation}
n_{4} \approx 1 - 2\frac{\omega_{\rm pe}^2<u>}{\omega^2} \, \theta_{\rm b}^2,
\label{n4ll}
\end{equation}
in agreement with~\citet{BGI88, Melrose2}.

\begin{figure}
\centerline{\includegraphics[height=5cm,width=10cm]{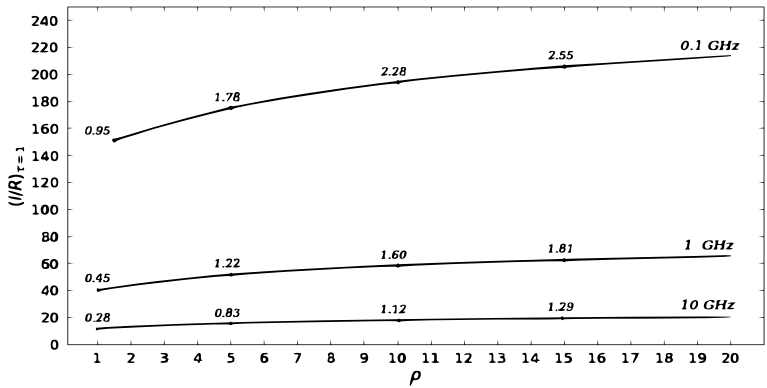}}
  \caption{The same as in Figure~\ref{fig04} for Alfv{\'e}n mode $j = 4$}
\label{fig05}
\end{figure}

Figure~\ref{fig05} shows the levels $l = r_{\rm d}$ at which the optical depth $\tau$ becomes equal to unity for the mode $j = 4$. Integration was carried out using general relations (\ref{go1})--(\ref{go2}) resulting in 
\begin{eqnarray}
\frac{{\rm d}r_{\perp}}{{\rm d} l} & = & \alpha, \\
\frac{{\rm d}\theta_{\perp}}{{\rm d} l} & = & \frac{3}{2} \, \frac{\theta_{\rm b}}{l}.
\end{eqnarray}
As before, we put the radiation radius $r_{0} = 3\,R$. The numbers again show the current ratios $\theta_{\rm b}/\theta^{\ast}$ at the level $\tau = 1$. As we can see, as for mode $j = 3$, even for $\rho = 1$ the wave damps at the heights significantly higher than the radiation level. On the other hand, for $\rho \sim 1$ the damping occurs at angles $\theta_{\rm b}$ even smaller than local $\theta^{\ast}(l)$. It should also be noted here that the approximation of the smallness of the second (imaginary) term in
 (\ref{3j4}) remains fulfilled.

\section{Alternative distribution function}
\label{sect:6}

\begin{figure}
\centering
\includegraphics[width=6cm]{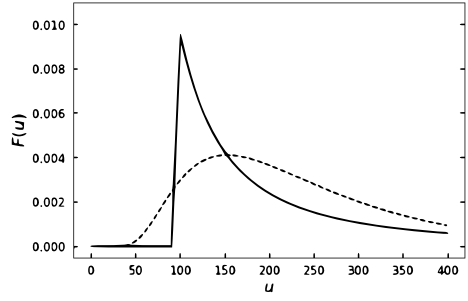}
\caption{Particle distribution function $F(u)$ (\ref{FUI}) in the laboratory frame, $ u_0=100, \, \Delta=10$. 
The dashed curve represents the J{\"u}ttner function with $\rho=4$ in the frame moving with the Lorentz factor $\gamma_s=210$.} 
\label{fig06}
\end{figure}

\begin{figure}
\centering
\includegraphics[width=6cm]{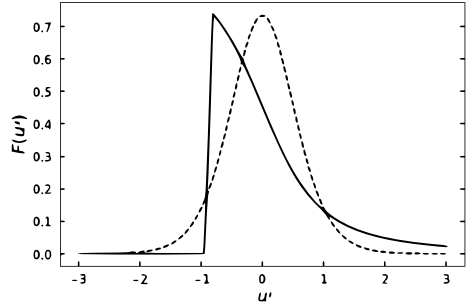}
\caption{Particle distribution function $F(u')$ (\ref{FUI}) in the rest frame where the mean velocity
is equal to zero. The Lorentz factor of motion of this frame is $\gamma_s = 210$. The dashed 
line shows the J{\"u}ttner function (\ref{Fpp}) with $\rho=4$.} 
\label{fig07}
\end{figure}

As was already stressed, J{\"u}ttner distribution function (\ref{FpJ}) may not accurately describe the real distribution function of particles flowing along open magnetic field lines in the pulsar magnetosphere. For this reason below we consider another particle distribution function in the laboratory frame (see Figures~\ref{fig06}, \ref{fig07})
\begin{equation}
  F(u) = \left\{
    \begin{array}{lll}
      0, & u<u_0-\Delta, \\[5pt]
      \frac{u-u_0+\Delta}{\Delta(u_0+\Delta/2)}, & u_0-\Delta<u<u_0, \\ [5pt]
      \frac{u_0^2}{(u_0+\Delta/2)} \, u^{-2},         & u>u_0.
    \end{array} \right.
\label{FUI}
\end{equation}
Accordingly, the derivative of the distribution function over $u$ is
\begin{equation}
  \frac{{\rm d}F(u)}{{\rm d}u} = \left\{
    \begin{array}{lll}
      0, & u<u_0-\Delta, \\[5pt]
      \frac{1}{\Delta(u_0+\Delta/2)}, & u_0-\Delta<u<u_0, \\ [5pt]
      -\frac{2u_0^2}{(u_0+\Delta/2)} \, u^{-3},         & u>u_0.
    \end{array} \right.
\label{FUI1}    
\end{equation}
In addition, we discuss in more detail the dispersion properties of the waves propagating in the pulsar magnetosphere. Finally, here we will not limit ourselves to small angles $\theta_{\rm b} \ll 1$.

Remind that the power-law dependence of the distribution function $F(u)$ (\ref{FUI}) on the momentum $u$ in the form $F(u) \propto u^{-2}$ for $u > u_0$ follows from numerical calculations of cascade generation of electron-positron plasma in the polar region of the pulsar magnetosphere first performed by~\citet{DH82} and subsequently confirmed by~\citet{GI85}. The characteristic value of $u$ is $u_0 \simeq 10^2$, and $\Delta \ll u_0$. The value of $\Delta$ defines the region of arising of the distribution function from zero to its maximum value at $u = u_0$. Finally, it is necessary to stress that the Lorentz boost transformation to the plasma rest frame ($<v> = 0$) corresponds to $\gamma_{\rm s} \approx 2 u_{0}$ (see Figure~\ref{fig07}). 

Since in the laboratory reference frame $u \gg 1$, we can put $\beta = 1 - 1/(2u^{2})$. As a result, dispersion equation (\ref{DDD}) can be rewritten as
\begin{equation}
1-n^2+\frac{1+n\cos\theta}{n\cos\theta}\frac{\omega_{\rm pe}^2}{\omega^2}\int\left[
1 + \frac{n\cos\theta}{2u^2(1-n\cos\theta)}\right]^{-1}\frac{{\rm d}F}{{\rm d}u} {\rm d}u = 0.
\label{DDDI}
\end{equation}

If 
\begin{equation}
|n\cos\theta_{\rm b}-1| > \frac{n\cos\theta_{\rm b}}{2(u_0 - \Delta)^2}
\end{equation}
(which is possible for $A_{\rm p} \gg 1$), the denominator in the integrand does not turn to zero (there is no resonance), and we can go from integration over the derivative of the distribution function to integration over the distribution function itself
\begin{eqnarray}
\int \left[
1 + \frac{n\cos\theta}{2u^2(1-n\cos\theta)}\right]^{-1}\frac{{\rm d}F}{{\rm d}u} {\rm d}u 
= - <\frac{n\cos\theta}{u^3(1-n\cos\theta)}
\left[
1 + \frac{n\cos\theta}{2u^2(1-n\cos\theta)}\right]^{-2}>.
\end{eqnarray}
Neglecting now the small term in straight brackets, we obtain for the dispersion equation
\begin{equation}
1-n^2-\frac{(1+n\cos\theta_{\rm b})}{1-n\cos\theta_{\rm b}} a_{\rm p}^{\prime} = 0,
\end{equation}
where now $a_{\rm p}^{\prime} = <\omega_{\rm pe}^2/(u^3\omega^2)> \approx a_{\rm p}$. For the distribution function $F(u)$ (\ref{FUI}), we obtain for $\Delta \ll u_{0}$
\begin{equation}
a_{\rm p}^{\prime} = \frac{1}{4} \, \frac{\omega_{\rm pe}^2}{u_{0}^{3}\omega^2}.
\end{equation}

Thus, for $u \gg 1$ dispersion equation (\ref{DDDI}) is reduced to a cubic equation
\begin{equation}
n^3 - \frac{1}{\cos\theta_{\rm b}} n^2 - (1+a_{\rm p}^{\prime})n + \frac{(1-a_{\rm p}^{\prime})}{\cos\theta_{\rm b}} = 0
\end{equation}
describing three electromagnetic modes propagating in the magnetosphere. Solving the cubic equation, we have
\begin{eqnarray}
n_2 & = & -\frac{(1-\sqrt{3}\,\mathrm{i})}{6}XZ^{-1/3}-\frac{(1+\sqrt{3}\,\mathrm{i})}{6}Z^{1/3}+\frac{1}{3\cos\theta_{\rm b}}, \\
n_3 & = & \frac{1}{3}(XZ^{-1/3}+Z^{1/3})+\frac{1}{3\cos\theta_{\rm b}}, \\
n_5 & = & -\frac{(1+\sqrt{3}\,\mathrm{i})}{6}XZ^{-1/3}-\frac{(1-\sqrt{3}\,\mathrm{i})}{6}Z^{1/3}+\frac{1}{3\cos\theta_{\rm b}},
\end{eqnarray}
where
\begin{eqnarray}
X & = & 3(1 + a_{\rm p}^{\prime})+\cos^{-2}\theta_{\rm b}, \\ 
Y & = & \cos^{-3}\theta_{\rm b} + 9(2a_{\rm p}^{\prime}-1)\cos^{-1}\theta_{\rm b}, \\ 
Z & = & (Y^2-X^{3})^{1/2}+Y.
\end{eqnarray}
Two roots, $n_2$ and $n_3$, which are already known by us, correspond to waves propagate in the positive direction ($n > 0$), and one extra mode $j = 5$ propagates in the negative direction ($n < 0$).

\begin{figure}
\centering
\includegraphics[width=6cm]{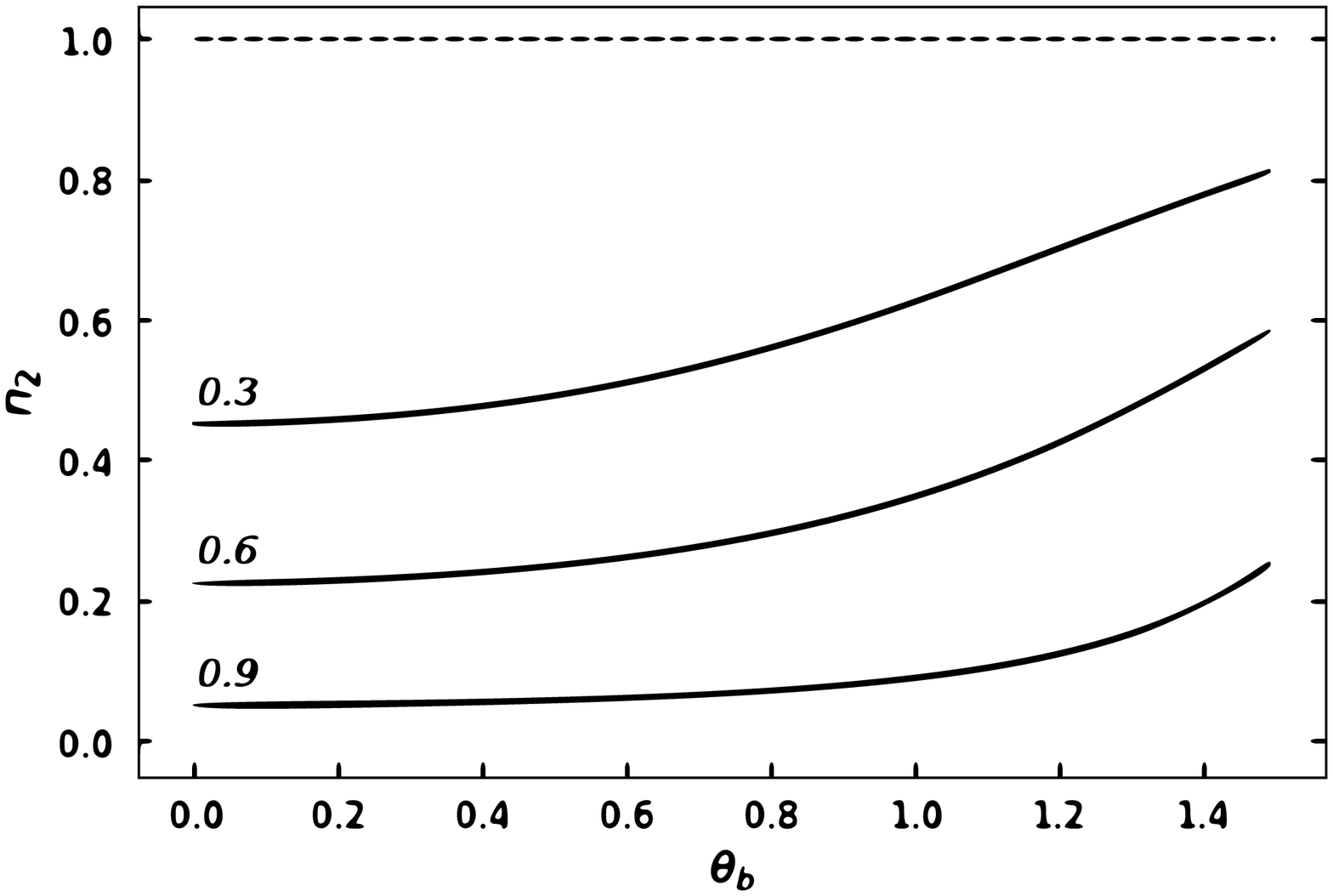}
\caption{Refractive index $n_{2}$ for mode $j = 2$. The parameter $a_{\rm p}^{\prime} = <\omega_p^2/u^{3}\omega^2> = 0.3$ (upper curve), $a_{\rm p}^{\prime} = 0.6$ (middle curve), and $a_{\rm p}^{\prime} = 0.9$ (lower curve). Dashed line corresponds to $n=1$.} 
\label{fig08}
\end{figure}

\begin{figure}
\centering
\includegraphics[width=6cm]{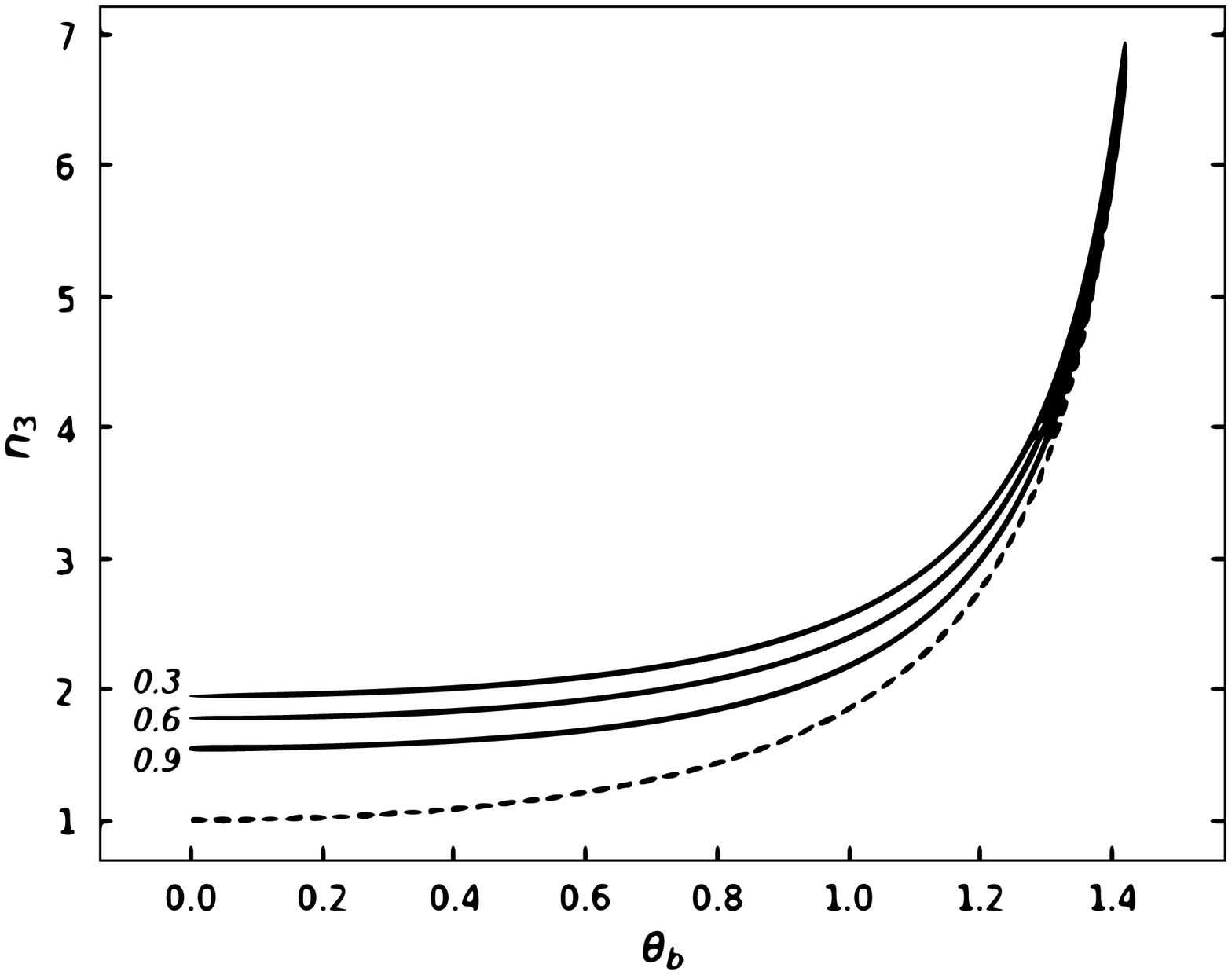}
\caption{Refractive index $n_{3}$ for mode $j = 3$. The parameter $a_{\rm p}^{\prime} = <\omega_p^2/u^{3}\omega^2> = 0.3$ (upper curve), $a_{\rm p}^{\prime} = 0.6$ (middle curve), and $a_{\rm p}^{\prime} = 0.9$ (lower curve). Dashed line corresponds to $n=1/\cos\theta_{\rm b}$.} 
\label{fig09}
\end{figure}

\begin{figure}
\centering
\includegraphics[width=6cm]{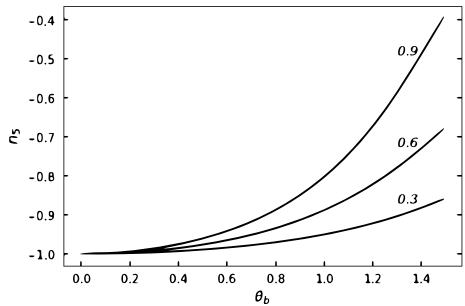}
\caption{Refractive index $n_{5}$ for mode $j = 5$. 
The parameter $a_{\rm p}^{\prime} = <\omega_p^2/u^3\omega^2> = 0.3$ (lower curve), $a_{\rm p}^{\prime} = 0.6$ (middle curve), and $a_{\rm p}^{\prime} = 0.9$ (upper curve).} 
\label{fig10}
\end{figure}

The dependencies of $n_2$, $n_3$, and $n_5$ on the angle $\theta_{\rm b}$  are presented in Figure~\ref{fig08}--\ref{fig10}. As was already shown, for $\theta_{\rm b} = 0$ ($\cos\theta_{\rm b} = 1$) we have $n_{2,3} = 1 \pm (a_{\rm p}^{\prime})^{1/2}$. But for \mbox{$\theta_{\rm b} = \upi/2$}, \mbox{($\cos\theta_{\rm b} = 0$)} we obtain $n_2 = (1-a_{\rm p}^{\prime})^{1/2}$ and $n_3 \to 1/\cos\theta$, superluminal mode \mbox{$j = 2$} transforming into a purely transverse wave. As to the wave $j = 5$, which was not mentioned earlier, it is a wave propagating in the negative direction in the laboratory reference frame, and is purely transverse: $n_{5} = -1$ at $\theta_{\rm b} = 0$ and $n_{5} = - (1 - a_{\rm p}^{\prime})^{1/2}$ at $\theta_{\rm b} = \upi/2$.
 %Modes $j = 2$ and $j = 3$ under longitudinal propagation along the magnetic field are electrostatic oscillations, $(n-1)^2 = <(\omega_{\rm pe}^2/\gamma^{-3}\omega^2)>$.
 
Returning now to the question of wave attenuation, we note that for $\Delta \ll u_{0}$ the attenuation of the plasma-Alfv{\'e}n wave $j = 3$, as for a cold plasma, begins only at angles $\theta_{\rm b} > \theta^{\ast}$. As for attenuation of the Alfv{\'e}n mode $j = 4$ ($n\cos\theta \simeq 1$), for this mode the second term in square brackets in (\ref{DDDI}) cannot be neglected. As a result, the integral becomes equal to
\begin{equation}
\int\left[
1 + \frac{n\cos\theta}{2u^2(1-n\cos\theta)}\right]^{-1}\frac{{\rm d}F}{{\rm d}u} {\rm d}u = \int\frac{u^2}{u^2-u_r^2-i0}\frac{{\rm d}F}{{\rm d}u} \, {\rm d}u.
\end{equation}
Here the quantity $u_r$ is the resonant momentum 
\begin{equation}
u_r^2 = \frac{n\cos\theta_{\rm b}}{2(n\cos\theta_{\rm b}-1)}, 
\end{equation}
and we added a small imaginary part in the denominator to take into account the Landau bypass rule for the pole $(\omega \to \omega + \mathrm{i} 0)$, i.e. for the refractive index $n \to n-\mathrm{i}0$. Thus, the dispersion equation takes the form
\begin{equation}
1 - n^2 + \frac{\omega_{\rm pe}^2}{\omega^2}\frac{1+n\cos\theta_{\rm b}}{n\cos\theta_{\rm b}}\left[\dashint \frac{u^2}{u^2-u_r^2}\frac{{\rm d}F}{{\rm d}u} \, {\rm d}u
+\mathrm{i}\frac{\upi u_r}{2}\frac{{\rm d}F}{{\rm d}u}\bigg|_{u=u_r}\right] = 0.
\end{equation}
Substituting now the derivative ${\rm d}F/{\rm d}u$ from (\ref{FUI1}) and again introducing the notation
\begin{equation}
x = n\cos\theta_{\rm b}-1,
\end{equation}
we obtain
\begin{equation}
\tan^2\theta_{\rm b} +\frac{2x}{\cos^2\theta_{\rm b}} - \frac{4\omega_{\rm pe}^2u_{0}}{\omega^2}\left[\log\left(1-\frac{1}{2xu_{0}^2}\right) - \mathrm{i}\upi\right]x = 0.
\end{equation}
For not very small angles $\theta_{\rm b} > u_0^{-2} \simeq 10^{-4} $ and the condition $u_0^2 \,  a_{\rm p}^{\prime} \gg 1$ we get
\begin{equation}
x = \mathrm{i}\frac{\tan^2\theta_{\rm b}}{4\upi}\frac{\omega^2}{u_{0}\omega_{\rm pe}^2}.
\end{equation}
Thus, the imaginary part of the refractive index of the Alfv{\'e}n mode is 
\begin{equation}
{\rm Im} (n_4 \cos\theta_{\rm b}) = \frac{\tan^2\theta_{\rm b}}{4\upi} \frac{\omega^2}{u_0\omega_{\rm pe}^2}\simeq \frac{\tan^2\theta_{\rm b}}{2\upi}A_{\rm p}^{-1},
\end{equation}
which corresponds to the attenuation of the wave during its propagation in the magnetosphere.

\begin{figure}
\centering
\includegraphics[width=6cm]{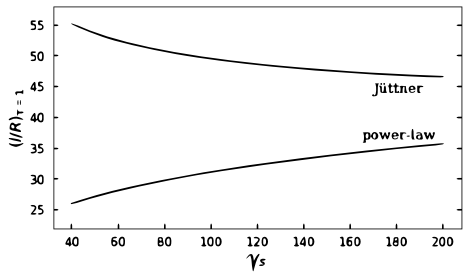}
\caption{Comparison of levels $l = r_{\rm d}$ at which the optical depth $\tau = 1$ (\ref{tau}) depending on parameter $\gamma_{\rm s}$ for both J{\"u}ttner ($\rho = 4$, $\gamma_{\rm s}$) and power-law ($u_0 = \gamma_{\rm s}/2$) distribution functions.} 
\label{fig11}
\end{figure}

On Figure~\ref{fig11} we show how the levels $l = r_{\rm d}$ at which the optical depth $\tau = 1$ (\ref{tau}) depend on parameter $\gamma_{\rm s}$ for both J{\"u}ttner ($\rho = 4$, $\gamma_{\rm s}$) and power-law ($u_0 = \gamma_{\rm s}/2$) distribution functions. As we see, these levels differ not so drastically. This is due to the fact that in both cases the attenuation is determined by resonant particles whose energy is not much higher than the average particle energy.

\section{Conclusion}

Thus, it was shown that thermal kinetic effects change not drastically the dispersion relations for all four waves propagating outwards in the pulsar magnetosphere. In particular, for superluminal O-mode $j = 2$ ''hydrodynamical'' relation (\ref{nj2}) gives good enough expression for refractive index $n_{2}$. We intend to conduct a detailed analysis of thermal effects on the observed average profiles of radio pulsars in a separate paper.

As to two subluminal modes $j = 3$ and $j = 4$, for them the kinetic effects result in more effective damping already at the angles $\theta_{\rm b} \sim \theta^{\ast}$. In particular, it was shown that the attenuation of a plasma-Alfv{\'e}n wave $j = 3$ substantially depends on the low-energy part of the spectrum of the outflowing particles: in the absence of a low-energy tail (see Sect.~\ref{sect:6}), attenuation occurs at the same distances from the star as for cold plasma. On the other hand, the presence of a high-energy power-law tail does not alter  significantly the attenuation of the Alfv{\'e}n wave $j = 4$ in comparison to the thermal spectrum. Recall that this not so important because these modes cannot escape the pulsar magnetosphere as at large distance from the neutron star they propagate along magnetic field lines.

In conclusion, let us comment three statements which was already mentioned above. First, it is necessary to stress that we confirmed once again that at zero angle $\theta_{\rm b}$ in the laboratory (pulsar) reference frame there are two branches of the same plasma wave propagating from a neutron star. But in the plasma rest frame both modes propagate inwards to the star surface, one of them ($j = 2$) being superluminal and the other ($j = 3$) subluminal. Second, it should be noted that our results do not contradict the conclusions obtained by~\citet{Melrose2}. Indeed, as was shown in this paper, in the region of phase velocities corresponding to $\gamma_{\phi} < 100$ (i.e., precisely in the region $A_{\rm p} > 1$), the results for plasma modes $j = 2, 3$ are actually identical. Thus, in our opinion, the question on the number of waves can be considered closed. Finally, recall that in the theory of radio emission proposed by~\citet{BGI88}, the key role is played by the instability of the Alfv{\'e}n mode $j = 4$ in the angle range $\theta_{\rm b} < \theta^{\ast}$ which is connected with the curvature of magnetic field lines. As was shown above, the inclusion of thermal effects into consideration does not lead to strong attenuation of the Alfv{\'e}n mode in this region. Thus, thermal effects do not violate significantly this key point of the theory.

\section*{Acknowledgements} 

We thank Sasha Philippov for useful discussions. This work was supported by Russian Foundation for Basic Research (Grant no. 17-02-00788).

\vspace{1cm}

\bibliographystyle{jpp}

\bibliography{BGIwaves}

\begin{thebibliography}{54}
\expandafter\ifx\csname natexlab\endcsname\relax\def\natexlab#1{#1}\fi
\def\au#1{#1} \def\ed#1{#1} \def\yr#1{#1}\def\at#1{#1}\def\jt#1{\textit{#1}}
  \def\bt#1{#1}\def\bvol#1{\textbf{#1}} \def\vol#1{#1} \def\pg#1{#1}
  \def\publ#1{#1}\def\arxiv#1{#1}\def\org#1{#1}\def\st#1{\textit{#1}}

\bibitem[{Andrianov} \& {Beskin}(2010)]{AB2010}
{\sc \au{{Andrianov}, A.~S.} \& \au{{Beskin}, V.~S.}} \yr{2010}  \at{{Limiting
  polarization effect - a key link in investigating the mean profiles of radio
  pulsars}}.  \jt{Astronomy Letters}  \bvol{36},  \pg{248--259}.

\bibitem[Arendt \& Eilek(2002)]{AE2002}
{\sc \au{Arendt, P.~N.} \& \au{Eilek, J.~A.}} \yr{2002}  \at{{Pair Creation in
  the Pulsar Magnetosphere}}.  \jt{Astrophys. J.}  \bvol{581},  \pg{451--469}.

\bibitem[Asseo {\em et~al.\/}(1983)Asseo, Pellat \& Sol]{APS83}
{\sc \au{Asseo, E.}, \au{Pellat, R.} \& \au{Sol, H.}} \yr{1983}  \at{{Radiative
  or two-stream instability as a source for pulsar radio emission}}.
  \jt{Astrophys. J.}  \bvol{266},  \pg{201--214}.

\bibitem[Asseo {\em et~al.\/}(1990)Asseo, Pelletier \& Sol]{APS90}
{\sc \au{Asseo, E.}, \au{Pelletier, G.} \& \au{Sol, H.}} \yr{1990}  \at{{A
  non-linear radio pulsar emission mechanism}}.  \jt{Mon. Not. R. Astron. Soc.}
   \bvol{247},  \pg{529--548}.

\bibitem[{Barnard} \& {Arons}(1986)]{BArons1986}
{\sc \au{{Barnard}, J.~J.} \& \au{{Arons}, J.}} \yr{1986}  \at{{Wave
  propagation in pulsar magnetospheres - Refraction of rays in the open flux
  zone}}.  \jt{Astrophys. J.}  \bvol{302},  \pg{138--162}.

\bibitem[Benford \& Buschauer(1977)]{BB77}
{\sc \au{Benford, G.} \& \au{Buschauer, R.}} \yr{1977}  \at{{Coherent pulsar
  radio radiation by antenna mechanisms: general theory}}.  \jt{Mon. Not. R.
  Astron. Soc.}  \bvol{179},  \pg{189--207}.

\bibitem[{Beskin} {\em et~al.\/}(1993){Beskin}, {Gurevich} \& {Istomin}]{BGI93}
{\sc \au{{Beskin}, V.}, \au{{Gurevich}, A.} \& \au{{Istomin}, Y.}} \yr{1993}
  {\em {Physics of the Pulsar Magnetosphere}\/}.  \publ{Cambridge University
  Press}.

\bibitem[{Beskin} {\em et~al.\/}(1988){Beskin}, {Gurevich} \& {Istomin}]{BGI88}
{\sc \au{{Beskin}, V.~S.}, \au{{Gurevich}, A.~V.} \& \au{{Istomin}, I.~N.}}
  \yr{1988}  \at{{Theory of the radio emission of pulsars}}.  \jt{Astrophys.
  Space Sci.}  \bvol{146},  \pg{205--281}.

\bibitem[{Beskin} \& {Philippov}(2012)]{BPh2012}
{\sc \au{{Beskin}, V.~S.} \& \au{{Philippov}, A.~A.}} \yr{2012}  \at{{On the
  mean profiles of radio pulsars - I. Theory of propagation effects}}.
  \jt{Mon. Not. R. Astron. Soc.}  \bvol{425},  \pg{814--840},  \arxiv{arXiv:
  1107.3775}.

\bibitem[Blandford(1975)]{Blandford75}
{\sc \au{Blandford, R.~D.}} \yr{1975}  \at{{Amplification of radiation by
  relativistic particles in a strong magnetic field}}.  \jt{Mon. Not. R.
  Astron. Soc.}  \bvol{179},  \pg{551--557}.

\bibitem[{Blandford} \& {Scharlemann}(1976)]{Blandford76}
{\sc \au{{Blandford}, R.~D.} \& \au{{Scharlemann}, E.~T.}} \yr{1976}  \at{{On
  the scattering and absorption of electromagnetic radiation within pulsar
  magnetospheres}}.  \jt{Mon. Not. R. Astron. Soc.}  \bvol{174},  \pg{59--85}.

\bibitem[{Blaskiewicz} {\em et~al.\/}(1991){Blaskiewicz}, {Cordes} \&
  {Wasserman}]{BCW91}
{\sc \au{{Blaskiewicz}, M.}, \au{{Cordes}, J.~M.} \& \au{{Wasserman}, I.}}
  \yr{1991}  \at{{A relativistic model of pulsar polarization}}.
  \jt{Astrophys. J.}  \bvol{370},  \pg{643--669}.

\bibitem[{Daugherty} \& {Harding}(1982)]{DH82}
{\sc \au{{Daugherty}, J.~K.} \& \au{{Harding}, A.~K.}} \yr{1982}
  \at{{Electromagnetic cascades in pulsars}}.  \jt{Astrophys. J.}  \bvol{252},
  \pg{337--347}.

\bibitem[{Dyks}(2008)]{Dyks2008}
{\sc \au{{Dyks}, J.}} \yr{2008}  \at{{Altitude-dependent polarization in radio
  pulsars}}.  \jt{Mon. Not. R. Astron. Soc.}  \bvol{391},  \pg{859--868},
  \arxiv{arXiv: 0806.0554}.

\bibitem[Gedalin {\em et~al.\/}(1999)Gedalin, Gruman \& Melrose]{GGM2002}
{\sc \au{Gedalin, M.}, \au{Gruman, E.} \& \au{Melrose, D.~B.}} \yr{1999}
  \at{{Mechanism of pulsar radio emission}}.  \jt{Mon. Not. R. Astron. Soc.}
  \bvol{337},  \pg{422--430}.

\bibitem[{Ginzburg} \& {Zheleznyakov}(1975)]{GZh}
{\sc \au{{Ginzburg}, V.~L.} \& \au{{Zheleznyakov}, V.}} \yr{1975}  \at{On the
  pulsar emission mechanisms}.  \jt{Ann. Rev. Astron. Astrophys.}  \bvol{13},
  \pg{511--535}.

\bibitem[Goldreich \& Keeley(1971)]{GK71}
{\sc \au{Goldreich, P.} \& \au{Keeley, D.~A.}} \yr{1971}  \at{{Coherent
  Synchrotron Radiation}}.  \jt{J. Plasma Phys.}  \bvol{170},  \pg{463--478}.

\bibitem[Gurevich \& Istomin(1985)]{GI85}
{\sc \au{Gurevich, A.~V.} \& \au{Istomin, I.~N.}} \yr{1985}  \at{{Generation of
  electron-positron plasma in a pulsar's magnetosphere}}.  \jt{Sov. Phys. JETP}
   \bvol{62},  \pg{1--11}.

\bibitem[{Hakobyan} {\em et~al.\/}(2017){Hakobyan}, {Beskin} \&
  {Philippov}]{HBP2017}
{\sc \au{{Hakobyan}, H.~L.}, \au{{Beskin}, V.~S.} \& \au{{Philippov}, A.~A.}}
  \yr{2017}  \at{{On the mean profiles of radio pulsars II: Reconstruction of
  complex pulsar light-curves and other new propagation effects}}.  \jt{Mon.
  Not. R. Astron. Soc.}  \bvol{469},  \pg{2704--2719}.

\bibitem[{Han} {\em et~al.\/}(1998){Han}, {Manchester}, {Xu} \& {Qiao}]{Han98}
{\sc \au{{Han}, J.~L.}, \au{{Manchester}, R.~N.}, \au{{Xu}, R.~X.} \&
  \au{{Qiao}, G.~J.}} \yr{1998}  \at{{Circular polarization in pulsar
  integrated profiles}}.  \jt{Mon. Not. R. Astron. Soc.}  \bvol{300},
  \pg{373--387},  \arxiv{arXiv: astro-ph/9806021}.

\bibitem[Hardee \& Rose(1978)]{HR78}
{\sc \au{Hardee, P.~E.} \& \au{Rose, W.~K.}} \yr{1978}  \at{Wave production in
  an ultrarelativistic electron-positron plasma}.  \jt{Astrophys. J.}
  \bvol{219},  \pg{274--287}.

\bibitem[Kazbegi {\em et~al.\/}(1991)Kazbegi, Machabeli \& Melikidze]{KMM91}
{\sc \au{Kazbegi, A.~Z.}, \au{Machabeli, G.~Z.} \& \au{Melikidze, G.~I.}}
  \yr{1991}  \at{{On the circular polarization in pulsar emission}}.  \jt{Mon.
  Not. R. Astron. Soc.}  \bvol{253},  \pg{377--387}.

\bibitem[Larroche \& Pellat(1987)]{LP87}
{\sc \au{Larroche, O.} \& \au{Pellat, R.}} \yr{1987}  \at{{Curvature
  instability of relativistic particle beams}}.  \jt{Phys. Rev. Lett.}
  \bvol{59},  \pg{1104--1107}.

\bibitem[Lominadze \& Mikhailovskii(1979)]{LM79}
{\sc \au{Lominadze, D.~G.} \& \au{Mikhailovskii, A.~B.}} \yr{1979}
  \at{{Longitudinal waves and the beam instability in a relativistic plasma}}.
  \jt{Sov. Phys. JETP}  \bvol{49},  \pg{483--489}.

\bibitem[Lominadze {\em et~al.\/}(1979)Lominadze, Mikhailovskii \&
  Sagdeev]{LMS79}
{\sc \au{Lominadze, D.~G.}, \au{Mikhailovskii, A.~B.} \& \au{Sagdeev, R.~Z.}}
  \yr{1979}  \at{{Langmuir turbulence of a relativistic plasma in a strong
  magnetic field}}.  \jt{Sov. Phys. JETP}  \bvol{50},  \pg{927--932}.

\bibitem[Lominadze {\em et~al.\/}(1983)Lominadze, Machabeli \& Usov]{LMU83}
{\sc \au{Lominadze, J.~G.}, \au{Machabeli, G.~Z.} \& \au{Usov, V.~V.}}
  \yr{1983}  \at{Theory of np:0532 pulsar radiation and the nature of the
  activity of the crab nebula}.  \jt{Astrophys. Space Sci.}  \bvol{90},
  \pg{19--43}.

\bibitem[{Lorimer} \& {Kramer}(2012)]{L&K}
{\sc \au{{Lorimer}, D.~R.} \& \au{{Kramer}, M.}} \yr{2012} {\em {Handbook of
  Pulsar Astronomy}\/}.  \publ{Cambridge University Press}.

\bibitem[{Lyne} \& {Graham-Smith}(2012)]{L&GS}
{\sc \au{{Lyne}, A.} \& \au{{Graham-Smith}, F.}} \yr{2012} {\em {Pulsar
  Astronomy}\/}.  \publ{Cambridge University Press}.

\bibitem[{Lyubarskii} \& {Petrova}(1998)]{LP98}
{\sc \au{{Lyubarskii}, Y.~E.} \& \au{{Petrova}, S.~A.}} \yr{1998}
  \at{{Refraction of radio waves in pulsar magnetospheres}}.  \jt{Astron.
  Astrophys.}  \bvol{333},  \pg{181--187}.

\bibitem[{Lyubarskii} \& {Petrova}(2000)]{LP00}
{\sc \au{{Lyubarskii}, Y.~E.} \& \au{{Petrova}, S.~A.}} \yr{2000}
  \at{{Propagation effects in pulsar magnetospheres}}.  \jt{Astron. Astrophys.}
   \bvol{355},  \pg{1168--1180}.

\bibitem[Lyubarsky(2008)]{L2008}
{\sc \au{Lyubarsky, Y.}} \yr{2008}  \at{{Pulsar emission mechanisms}}.  \bt{In
  {\em 40 years of pulsars: Millisecond Pulsars, Magnetars and More.\/} (ed.
  \ed{A.~Cumming C.~G.~Bassa, Z.~Wang \& V.M. Kaspi})},  \st{AIP Conference
  Proceedings},  \vol{vol. 983},  \pg{pp. 29--37}.

\bibitem[Lyutikov(1999)]{L99}
{\sc \au{Lyutikov, M.}} \yr{1999}  \at{{Beam instabilities in a magnetized pair
  plasma}}.  \jt{J. Plasma Phys.}  \bvol{62},  \pg{65--86}.

\bibitem[Lyutikov {\em et~al.\/}(1999)Lyutikov, Blandford \& Machabeli]{LBM99}
{\sc \au{Lyutikov, M.}, \au{Blandford, R.} \& \au{Machabeli, G.}} \yr{1999}
  \at{{On the nature of pulsar radio emission }}.  \jt{Mon. Not. R. Astron.
  Soc.}  \bvol{305},  \pg{338--352}.

\bibitem[{Manchester} \& {Taylor}(1977)]{M&T}
{\sc \au{{Manchester}, R.} \& \au{{Taylor}, J.}} \yr{1977} {\em {Pulsars}\/}.
  \publ{Freeman, San Francisco}.

\bibitem[Medin \& Lai(2010)]{ML2010}
{\sc \au{Medin, Z.} \& \au{Lai, D.}} \yr{2010}  \at{{Pair cascades in the
  magnetospheres of strongly magnetized neutron stars}}.  \jt{Mon. Not. R.
  Astron. Soc.}  \bvol{406},  \pg{1379--1404}.

\bibitem[Melrose \& Gedalin(1999)]{MG99}
{\sc \au{Melrose, D.~B.} \& \au{Gedalin, M.~E.}} \yr{1999}  \at{{Relativistic
  Plasma Emission and Pulsar Radio Emission: A Critique}}.  \jt{Astrophys. J.}
  \bvol{521},  \pg{351--361}.

\bibitem[Melrose \& Rafat(2017)]{Melrose0}
{\sc \au{Melrose, D.~B.} \& \au{Rafat, M.~Z.}} \yr{2017}  \at{{Pulsar radio
  emission mechanism: Why no consensus?}}  \jt{J. Phys.: Conf. Series}
  \bvol{932},  \pg{012011}.

\bibitem[{Mestel}(1999)]{Mestel}
{\sc \au{{Mestel}, L.}} \yr{1999} {\em {Pulsars}\/}.  \publ{Clarendon, Oxford}.

\bibitem[{Michel}(1991)]{Michel}
{\sc \au{{Michel}, F.}} \yr{1991} {\em {Theory of pulsar magnetospheres}\/}.
  \publ{University of Chicago Press, Chicago}.

\bibitem[{Petrova}(2001)]{P2001}
{\sc \au{{Petrova}, S.~A.}} \yr{2001}  \at{{The effect of magnetospheric
  refraction on the morphology of pulsar profiles}}.  \jt{Mon. Not. R. Astron.
  Soc.}  \bvol{360},  \pg{592--602}.

\bibitem[{Philippov} {\em et~al.\/}(2019){Philippov}, {Uzdensky}, {Spitkovsky}
  \& {Cerutti}]{PhUSC}
{\sc \au{{Philippov}, A.}, \au{{Uzdensky}, D.}, \au{{Spitkovsky}, A.} \&
  \au{{Cerutti}, B.}} \yr{2019}  \at{Pulsar radio emission mechanism: Radio
  nanoshots as a low-frequency afterglow of relativistic magnetic
  reconnection}.  \jt{Astrophys. J.}  \bvol{876},  \pg{6--12}.

\bibitem[Rafat {\em et~al.\/}(2019{\natexlab{{\em a\/}}})Rafat, Melrose \&
  Mastrano]{Melrose1}
{\sc \au{Rafat, M.~Z.}, \au{Melrose, D.~B.} \& \au{Mastrano, A.}}
  \yr{2019{\natexlab{{\em a\/}}}}  \at{{Wave dispersion in pulsar plasma: 1.
  Plasma rest frame}}.  \jt{J. Plasma Phys.}  \bvol{85},  \pg{905850305}.

\bibitem[Rafat {\em et~al.\/}(2019{\natexlab{{\em b\/}}})Rafat, Melrose \&
  Mastrano]{Melrose2}
{\sc \au{Rafat, M.~Z.}, \au{Melrose, D.~B.} \& \au{Mastrano, A.}}
  \yr{2019{\natexlab{{\em b\/}}}}  \at{{Wave dispersion in pulsar plasma: 2.
  Pulsar frame}}.  \jt{J. Plasma Phys.}  \bvol{85},  \pg{905850311}.

\bibitem[Suvorov \& Chugunov(1975)]{SCh75}
{\sc \au{Suvorov, E.~V.} \& \au{Chugunov, I.~V.}} \yr{1975}
  \at{{Electromagnetic waves in a relativistic plasma with a strong magnetic
  field}}.  \jt{Astrophysics}  \bvol{11},  \pg{203--222}.

\bibitem[{Timokhin}(2010)]{T2010}
{\sc \au{{Timokhin}, A.~N.}} \yr{2010}  \at{{A model for nulling and mode
  changing in pulsars}}.  \jt{Mon. Not. R. Astron. Soc.}  \bvol{408},
  \pg{L41–L45}.

\bibitem[{Timokhin} \& {Harding}(2015)]{TH2015}
{\sc \au{{Timokhin}, A.~N.} \& \au{{Harding}, A.~K.}} \yr{2015}  \at{{On the
  Polar Cap Cascade Pair Multiplicity of Young Pulsars}}.  \jt{Astrophys. J.}
  \bvol{810},  \pg{144},  \arxiv{arXiv: 1504.02194}.

\bibitem[Ursov \& Usov(1988)]{UU88}
{\sc \au{Ursov, V.~N.} \& \au{Usov, V.~V.}} \yr{1988}  \at{{Plasma Flow
  Nonstationarity in Pulsar Magnetospheres and Two-Stream Instability}}.
  \jt{Astrophys. Space Sci.}  \bvol{140},  \pg{325--336}.

\bibitem[Usov(1987)]{U87}
{\sc \au{Usov, V.~V.}} \yr{1987}  \at{{On Two-Stream Instability in Pulsar
  Magnetospheres}}.  \jt{Astrophys. J.}  \bvol{320},  \pg{333--335}.

\bibitem[Usov(2006)]{U2006}
{\sc \au{Usov, V.~V.}} \yr{2006}  \at{{Radio emission theories of pulsars}}.
  \bt{In {\em On the Present and Future of Pulsar Astronomy\/} (ed. \ed{J.~Gil
  W.~Becker \& B.~Rudak})},  \st{Highlights of Astronomy},  \vol{vol. 218},
  \pg{p. 112}.

\bibitem[Wang {\em et~al.\/}(2010)Wang, Lai \& Han]{Wang1}
{\sc \au{Wang, C.}, \au{Lai, D.} \& \au{Han, J.}} \yr{2010}  \at{Polarization
  changes of pulsars due to wave propagation through magnetospheres}.  \jt{Mon.
  Not. R. Astron. Soc.}  \bvol{403},  \pg{569--588}.

\bibitem[Wang {\em et~al.\/}(2014)Wang, Wang \& Han]{Wang3}
{\sc \au{Wang, P.~F.}, \au{Wang, C.} \& \au{Han, J.~L.}} \yr{2014}
  \at{Polarized curvature radiation in pulsar magnetosphere}.  \jt{Mon. Not. R.
  Astron. Soc.}  \bvol{441},  \pg{1943--1953}.

\bibitem[Wang {\em et~al.\/}(2015)Wang, Wang \& Han]{Wang4}
{\sc \au{Wang, P.~F.}, \au{Wang, C.} \& \au{Han, J.~L.}} \yr{2015}  \at{On the
  frequency dependence of pulsar linear polarization}.  \jt{Mon. Not. R.
  Astron. Soc.}  \bvol{448},  \pg{771--780}.

\bibitem[{Weatherall}(1994)]{W94}
{\sc \au{{Weatherall}, J.~C.}} \yr{1994}  \at{Streaming instability in
  relativistically hot pulsar magnetospheres}.  \jt{Astrophys. J.}  \bvol{428},
   \pg{261–266}.

\bibitem[{Yuen} \& {Melrose}(2014)]{YuMelrose}
{\sc \au{{Yuen}, R.} \& \au{{Melrose}, D.~B.}} \yr{2014}  \at{{Visibility of
  Pulsar Emission: Motion of the Visible Point}}.  \jt{Publ. Astron. Soc.
  Australia}  \bvol{31},  \pg{e039}.

\end{thebibliography}

\end{document}